# Demonstrating a Service-Enhanced Retrieval System

**Philipp Schaer**
GESIS – Leibniz-Institute for the Social Sciences
Lennéstr. 30, 53113 Bonn, Germany
philipp.schaer@gesis.org

**Philipp Mayr**
GESIS – Leibniz-Institute for the Social Sciences
Lennéstr. 30, 53113 Bonn, Germany
philipp.mayr@gesis.org

**Peter Mutschke**
GESIS – Leibniz-Institute for the Social Sciences
Lennéstr. 30, 53113 Bonn, Germany
peter.mutschke@gesis.org

## ABSTRACT

This paper is a short description of an information retrieval system enhanced by three model driven retrieval services: (1) co-word analysis based query expansion, re-ranking via (2) Bradfordizing and (3) author centrality. The different services each favor quite other – but still relevant – documents than pure term-frequency based rankings. Each service can be interactively combined with each other to allow an iterative retrieval refinement.

## OVERVIEW

To overcome the limitations of metadata-driven Digital Libraries the Value-Added Services for Information Retrieval (IRM) project is doing research on an overall approach to use computational science models as enhanced search stratagems [Bates, 1990] within a scholarly retrieval environment, which might be implemented within scholarly information portals. We will propose an interactive demonstration system, which implements three different services to allow an iterative retrieval refinement in scholarly information systems: (1) a co-word analysis based query expansion, re-ranking via (2) Bradfordizing and (3) author centrality. Each of the three services provides a particular view to the information space that is quite different from traditional retrieval methods.

## SERVICES FOR RETRIEVAL ENHANCEMENT

All proposed models are implemented in a live information system using (1) the Solr search engine, (2) Grails Web framework and (3) Recommind Mindserver to demonstrate the general feasibility of the approaches. Solr is an open source search platform from the Apache Lucene project. The Mindserver is a text categorization tool, which was used to generate the query expansion terms. Both re-ranking mechanism Bradfordizing and author centrality are implemented as plugins to the open source web framework Grails, which is the glue to combine the different modules and to offer an interactive web-based prototype.

The system[1] allows the user to formulate his/her query, which could be interactively or automatically enriched by a Search Term Recommender (STR) that adds controlled descriptors form the corresponding document language. With this new query a search in the databases can be triggered. The result sets obtained can then be re-ranked using either Bradfordizing or author centrality.

### Databases

The prototype includes bibliographic metadata from six different scientific databases from three different scientific disciplines (social science, sport science and pedagogy). The databases differ in size, language, indexing depth and metadata schema. The databases included are:

- *SOLIS* (GESIS): 369,397 German and English literature references on social science
- *SPOLIT* (Bundesinstitut für Sportwissenschaft): 162,737 German literature references on sport science
- *FIS Bildung* (Deutsches Institut für Internationale Pädagogische Forschung): 697,528 German literature references on pedagogy
- *SA* (Cambridge Scientific Abstracts): Sociological Abstracts, 376,191 English literature references on social science
- *PEI* (Cambridge Scientific Abstracts): Physical Education Index, 127,754 English literature references on sport science
- *FES* (Friedrich-Ebert-Stiftung): Library catalogue, 487,559 German literature references

### Query Expansion by Search Term Recommendation

When using search in an information system a user has to come up with the "correct" terms to formulate his query. These terms have to match the terms used in the documents to get an appropriate result. This is long known as the language problem in IR (Petras 2006).

Work presented at:
*ASIST 2010,* October 22–27, 2010, Pittsburgh, PA, USA.

[1] http://www.gesis.org/beta/prototypen/irm

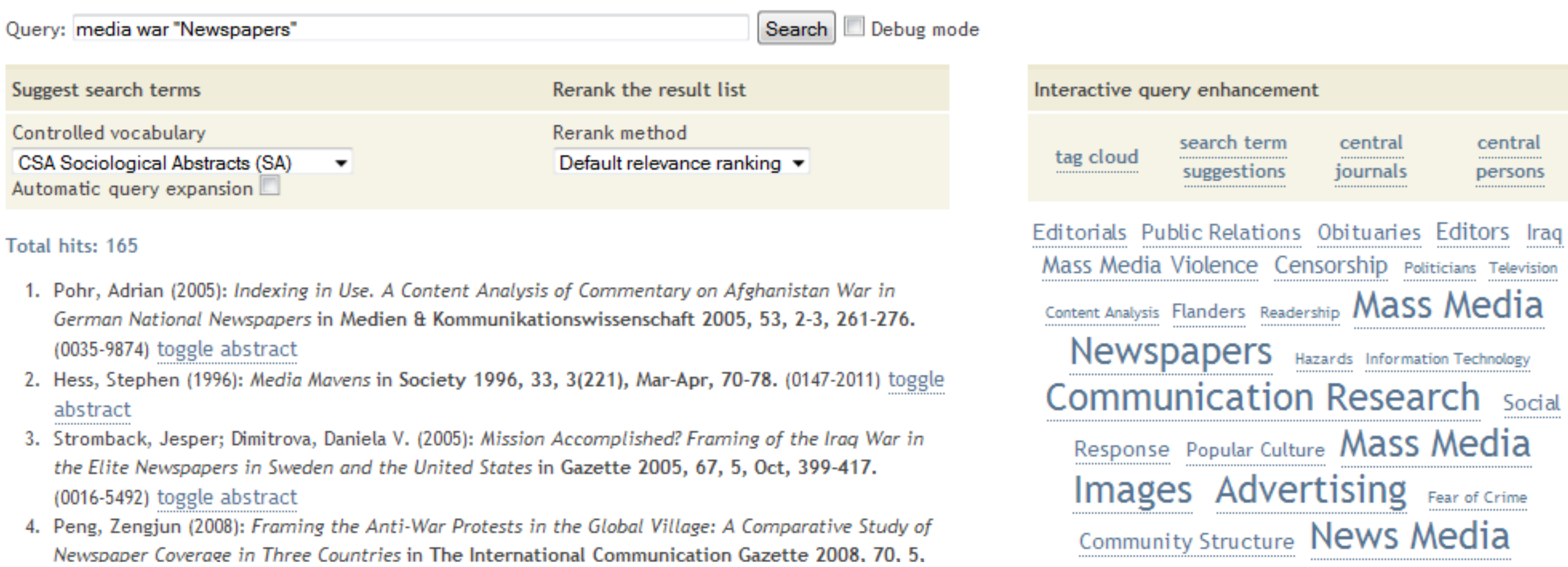

**Figure 1: Term suggestions on a query on "media war" – visualized in a term cloud. The other services are interactively available via tabbed browsing in the right column or as automatic services in the middle.**

The STR is based on statistical co-word analysis and builds associations between free terms (e.g. from title or abstract) and controlled terms (e.g. from a thesaurus). The software – which uses Support Vector Machines (SVM) and Probabilistic Latent Semantic Analysis (PLSA) – was trained with the six different databases included in the system to best match the specialized vocabularies used in the Social Sciences. The STR can be used as a Term Cloud, an ordered list of terms or as an automatic query expansion mechanism with six specialized vocabularies, matching the used databases.

### Re-Ranking by Bradfordizing

The Bradfordizing re-ranking service addresses the problem of oversized result sets by using a bibliometric method. Bradfordizing re-ranks a result set of journal articles according to the frequency of journals in the result set such that articles of core journals – journals that publish frequently on a topic – are ranked higher than articles from other journals (White 1981).

Our implementation of re-ranking by Bradfordizing is a simple approach, which is quickly implementable with Solr's build-in facet functionality. The only precondition for the application is the existence of qualitative metadata (ISSN in our case) to assure precise identification of the documents. In a first step the search results are filtered according to their ISSN. The next step aggregates all results with the same ISSN using facets. The journal with the highest ISSN facet count gets the top position in the result set; the second journal gets the next position, and so on. In the last step, each documents rank (given through Solr's internal ranking) is boosted by the frequency counts of the journals. This way all documents from the journal with the highest ISSN facet count are sorted to the top.

### Re-Ranking by Author Centrality

Author centrality is another way of re-ranking result sets. Here the concept of centrality of authors in a network is an approach for the problem of large and unstructured result sets. The model assumes that the relevancy of publications increases with the centrality of their authors in co-authorship networks. Since the model is based on a network analytical view it differs greatly from text-oriented ranking methods like *tf-idf*.

The re-ranking service computes a co-authorship network based on the result set to a specific query. Centrality of each single author in this network is calculated by applying the betweenness measure. The documents in the result set are then ranked according to the betweenness of their authors such that publications of very central authors are ranked higher in the result list.

## DEMONSTRATION REQUIREMENTS

The prototype is a web-based system, so the demonstration needs a permanent online connection. The demonstration itself will be conducted on a standard laptop, which will be brought in by us. Any other system with an Internet browser (Mozilla Firefox preferred) is working, too.

## ACKNOWLEDGMENTS

This work was funded by DFG, grant no. INST 658/6-1.

## REFERENCES

Bates, M. J. (1990). Where Should the Person Stop and the Information Search Interface Start? *Information Processing & Management*, 26 575-591.

Mayr, P., Mutschke, P., & Petras, V. (2008). Reducing semantic complexity in distributed digital libraries: Treatment of term vagueness and document re-ranking. *Library Review*, 57(3), 213-224.

Petras, V. (2006). *Translating Dialects in Search: Mapping between Specialized Languages of Discourse and Documentary Languages.* University of California, Berkley.

White, H. D. (1981). 'Bradfordizing' search output: how it would help online users. *Online Review, 5*(1), 47-54.